\documentclass[prx,notitlepage,twocolumn,superscriptaddress]{revtex4-1} % DS : only revtex4 works for me. MC: I re-enabled revtex4-1 in order to have better bibliography with links (without explicit url adress) 
\usepackage{amssymb}
\usepackage{bbold}
\usepackage{hyperref}  % DS : no pdftex option for me. I use lualatex.
\hypersetup{plainpages=false,colorlinks=true,linkcolor=blue, citecolor=blue, urlcolor=blue}

\usepackage{graphicx}
\usepackage{amsmath}
\usepackage{color}
\usepackage{float} 
\usepackage{multirow}
\usepackage{placeins}
\usepackage{xcolor}
\pagecolor{white}

\newcommand{\ket}[1]{\left| #1 \right\rangle} % for Dirac bras
\newcommand{\bra}[1]{\left\langle #1 \right|} % for Dirac kets

\begin{document}

\title{Fermi Arcs From Dynamical Variational Monte Carlo}
\author{P. Rosenberg}
\affiliation{Département de Physique, RQMP \& Institut Quantique, Université de Sherbrooke, Québec, Canada J1K 2R1}
\author{D. Sénéchal}
\affiliation{Département de Physique, RQMP \& Institut Quantique, Université de Sherbrooke, Québec, Canada J1K 2R1}
\author{A.-M. S. Tremblay}
\affiliation{Département de Physique, RQMP \& Institut Quantique, Université de Sherbrooke, Québec, Canada J1K 2R1}
\author{M. Charlebois}
\affiliation{Département de Physique, RQMP \& Institut Quantique, Université de Sherbrooke, Québec, Canada J1K 2R1}
\affiliation{Département de Chimie, Biochimie et Physique, Institut de Recherche sur l’Hydrogène, Université du Québec à Trois-Rivières, Trois-Rivières, Québec G9A 5H7, Canada}

\begin{abstract}
Variational Monte Carlo is a many-body numerical method that scales well with system size. It has been extended to study the Green function only recently by Charlebois and Imada (2020). Here we generalize the approach to systems with open boundary conditions in the absence of translational invariance. 
Removing these constraints permits the application of embedding techniques like Cluster perturbation theory (CPT). This allows us to solve an enduring problem in the physics of the pseudogap in cuprate high-temperature superconductors, namely the existence or absence of Fermi arcs in the one-band Hubbard model. 
We study the behavior of the Fermi surface and of the density of states as a function of 
hole doping for clusters of up to 64 sites, well beyond the reach of modern exact diagonalization solvers. We observe 
that the technique reliably captures the transition from a Mott insulator at half filling to a pseudogap, evidenced 
by the formation of Fermi arcs, and finally to a metallic state at large doping. The ability to treat large clusters with
quantum cluster methods helps to minimize potential finite size effects and enables the study of systems with long range orders, 
which will help extend the reach of these already powerful methods and provide important insights on the nature of various strongly 
correlated many-electron systems, including the high-T$_c$ cuprate superconductors.    
\end{abstract}

\maketitle

%===============================================================================
\section{Introduction}

Solving important problems in the field of correlated electrons often requires the development of new numerical methods.
Currently, the toolbox at our disposal includes; quantum Monte Carlo, capable of treating fairly large systems, but limited by the fermion sign problem to a rather restricted set of problems or parameters; exact diagonalization that is free of the sign problem, but limited to small systems, diagrammatic Monte Carlo  and variational methods, including iPEPS and DMRG~\cite{Qin_Schafer_Andergassen_Corboz_Gull_2022}.

Here, we develop and benchmark a numerical approach based on sign-problem-free variational Monte Carlo (VMC)~\cite{misawa_mvmcopen-source_2019-1, tahara_variational_2008-1, charlebois_single-particle_2020-1, mezzacapo_ground-state_2009, stella_strong_2011, sorella_wave_2005, zhao_variational_2017-1,nagy_variational_2019,Han2019, Choo2020,Pfau2020,Stokes2020}. We generalize the Green function calculation recently introduced in dynamical variational Monte Carlo (dVMC)~\cite{charlebois_single-particle_2020-1} in order to relax the constraints of periodic boundary conditions and translation invariance.
This generalized dVMC can treat clusters with open boundaries, scales well as a function of system size, does not require any analytic continuation and works for any interaction strength;
it can be incorporated into embedding approaches, 
 such as dynamical mean-field theory~\cite{georges_dynamical_1996-3,kotliar_electronic_2006-3,hansmann_motthubbard_2013} or its cluster extensions~\cite{hettler_nonlocal_1998,hettler_dynamical_2000-1,aryanpour_analysis_2002,maier_quantum_2005-3,LTP:2006,lichtenstein_antiferromagnetism_2000-4, kotliar_cellular_2001-3,maier_quantum_2005-3}, which are a powerful set of tools to study correlated many-electron systems. 

We combine our generalization of dVMC with a simple embedding method, cluster perturbation theory~\cite{gros_cluster_1993, senechal_spectral_2000,Senechal2002}, 
to answer the important question of the existence of Fermi arcs in paramagnetic state of the one-band Hubbard model, as it applies to high-$T_c$ cuprates. 

Let us explain this problem in the context of the ground state of cuprates in strong magnetic fields, where superconductivity is destroyed or has negligible effects. 
At half filling, these materials are antiferromagnetic Mott insulators. 
Upon increasing the hole concentration $p$ from zero, an itinerant antiferromagnet survives up to $p\sim 0.05$~\cite{kunisada_observation_2020}, followed by a pseudogap phase.
The pseudogap phase extends to roughly $p=p^*\sim0.19$~\cite{ding_spectroscopic_1996,Tallon_Loram_2001}, beyond which the material settles in a Fermi liquid state, at around $p=0.25$.
An interviening charge-density-wave state~\cite{Frano_Blanco-Canosa_Keimer_Birgeneau_2019} is often present within the pseudogap phase. Local-spin moments and glassy behavior are also observed in some compounds~\cite{Vinograd_Zhou_Mayaffre_Kramer_Ramakrishna_Reyes_Julien_2022}.

The pseudogap phase remains a puzzling phenomenon in cuprates. 
It is characterized by Fermi arcs~\cite{norman_destruction_1998, kanigel_evolution_2006, marshall_unconventional_1996-1} observed in angle-resolved photoemission spectroscopy (ARPES)~\cite{damascelli_angle-resolved_2003-2} and in Scanning Tunneling Microscopy~\cite{McElroy_Lee_Hoffman_Lang_Lee_Hudson_Eisaki_Uchida_Davis_2005}. 
The absence of a connected Fermi surface contradicts the intuition provided by band theory and apparently violates Luttinger's theorem.
Moreover, there is a drastic change in the carrier density from $p$ to $1+p$, as measured by Hall conductivity, when increasing $p$ through $p^*$~\cite{fang_fermi_2022,collignon_fermi-surface_2017,badoux_change_2016-2}.  

Many different phenomenological models predict arc-like structures~\cite{reber_origin_2012,scheurer2018, eberlein_fermi_2016-2,zhang_pseudogap_2020,singh_fermi_2022, chen_fermi_2021,verret_fermi_2022-1}.
However, when different phenomenological theories with different ingredients predict the same phenomenon, it becomes difficult to discriminate exactly which ingredient is responsible for that phenomenon.
A more reliable approach is to start from a simple microscopic model of interacting electrons, for instance the $t-J$ or Hubbard model, to see whether the phenomenon naturally emerges from such a minimal description.
This is the course of action followed here.

In order to study the Fermi surface, it is important to use an embedding approach such as Cluster perturbation theory (CPT)~\cite{gros_cluster_1993, senechal_spectral_2000,Senechal2002} or cluster-DMFT (c-DMFT)~\cite{lichtenstein_antiferromagnetism_2000-4, kotliar_cellular_2001-3,maier_quantum_2005-3}\footnote{The Dynamical Cluster Approximation (DCA), a cluster extension of DMFT defined in reciprocal space, is not well suited to study the Fermi surface because of the discontinuous nature of the self-energy in reciprocal space.}.
With these embedding approaches, it is possible~\cite{verret_fermi_2022-1} to recover the full $\mathbf{k}$-point dependence of the Fermi surface~\cite{senechal_cluster_2002, stanescu_fermi_2006, biroli_cluster_2002, biroli_cluster_2004, sakai_cluster-size_2012, verret_intrinsic_2019,verret_fermi_2022-1} using a \emph{periodization} scheme for the cluster Green function.
A number of studies have reported what looks like Fermi arcs using these techniques~\cite{stanescu_fermi_2006,senechal_hot_2004-1,Kancharla:2008}, with the caveat of low energy resolution caused by the small cluster size. 
In order to lift this caveat, larger clusters are necessary. The method presented in this work surmounts this significant technical hurdle.
We will show that this general dVMC approach provides strong evidence of Fermi arcs, and reproduces directly the ARPES Fermi surface across the whole pseudogap transition. In DCA, the destruction of the Fermi surface in the pseudogap is attributed to a momentum-selective Mott transition~\cite{Gull_Parcollet_Werner_Millis_2009,Gull_Ferrero_Parcollet_Georges_Millis_2010,Gull_Parcollet_Millis_2013} 

The remainder of the paper is organized as follows. 
We introduce the method in section~\ref{se::method}. 
In section~\ref{se::results} we look at the Fermi surface and density of states of the Hubbard model with parameter sets simulating underdoped hole-doped cuprates. 
Finally, we discuss the meaning of these results and potential further applications of the dVMC technique in section~\ref{se::discussion}. 
Details of the approach and benchmarking are provided in Appendices. 

%===============================================================================
\section{Method}
\label{se::method}
In this section we introduce the general dVMC technique,
a method to compute the Green function for strongly-correlated electrons.
This extends the recently developed dVMC method \cite{charlebois_single-particle_2020-1}
to treat systems with general boundary conditions, without assuming translational 
invariance. 

%-------------------------------------------------------------------------------
\subsection{VMC for the ground state}
\label{vmc}

The first step of the dVMC method is to compute the VMC ground state. We therefore begin with a
brief description of the VMC method as applied to lattice systems,
for which the occupation number basis is most convenient.

In VMC, the expectation value of any operator $\hat A$ is computed as
\begin{align}
\langle \hat A\rangle&=\frac{\langle\Omega|\hat A| \Omega\rangle}{\langle\Omega | \Omega\rangle}=\sum_{x} \frac{\langle\Omega|\hat A| x\rangle\langle x | \Omega\rangle}{\langle\Omega | \Omega\rangle}
\label{vmcSummary}
\\
&=\sum_{x} \rho(x) \frac{\langle\Omega|\hat A| x\rangle}{\langle\Omega | x\rangle}
\;\; \text{where} \;\;  \rho(x)=\frac{|\langle x | \Omega\rangle|^{2}}{\langle\Omega | \Omega\rangle}
\end{align}
where $| \Omega\rangle$ is the variational ground state and the sum $\sum_{x} |x\rangle\langle x|$ is over the complete set of all possible electronic configurations for the system. 
The number of possible configurations grows in a combinatorial fashion with the number of electrons and lattice sites, making it computationally
intractable to sum over every configuration for large systems. Instead, this sum is estimated via Monte Carlo sampling, which can achieve arbitrary
precision with a sufficient number of samples. 

The Monte Carlo estimate for the expectation value of any observable is given by
\begin{align}
\label{vmcSummary2} 
\langle \hat A\rangle_{\rm MC}
&= \frac{1}{N_{\rm MC}} \sum_{s}^{N_{\rm MC}} \frac{\langle\Omega|\hat A| s \rangle}{\langle\Omega | s \rangle}
\end{align}
where $N_{\text{MC}}$ is the number of Monte Carlo samples where each sample $s$ is a specific electronic configuration in real space, each of which is generated with probability $\rho(s)$ using a Metropolis algorithm.

VMC calculations typically begin with an ansatz for the ground state wavefunction.
Here we choose the representation employed in Refs~\cite{misawa_mvmcopen-source_2019-1, tahara_variational_2008-1, charlebois_single-particle_2020-1}:
\begin{align}
\label{psi_ground}
|\Omega\rangle&=\mathcal{P}_{\mathcal{G}} \mathcal{P}_{\mathcal{J}} %\mathcal{P}^{4}_{d-h} 
|\phi\rangle 
\\
|\phi\rangle&=\left(\sum_{i, j} f_{i j} \hat c_{i \uparrow}^{\dagger} \hat c_{j \downarrow}^{\dagger}\right)^{N_{e} / 2}|0\rangle 
\\ 
\mathcal{P}_{\mathcal{G}}&=\exp \left(\sum_{i} g_i \hat n_{i \uparrow} \hat n_{i \downarrow}\right) 
\\ 
\mathcal{P}_{\mathcal{J}}&=\exp \left(\sum_{i\ne j} v_{i j} \hat n_{i} \hat n_{j}\right)
\end{align}
where $|0\rangle$ is the vacuum state, $\hat c_{i \sigma}^{\dagger}$ is the creation operator for an electron of spin-$\sigma$ at site $i$, $\hat n_{i \sigma}= \hat c_{i \sigma}^{\dagger}\hat c_{i \sigma}$, and $N_e$ is the total number of electrons in the ground state (an even number). The parameters $f_{i j}$, $g_i$ and $v_{i j}$ are variational degrees of freedom whose values are determined via minimization of the ground state energy $\Omega = \langle\Omega|\hat H| \Omega\rangle / \langle\Omega| \Omega\rangle$. 

%-------------------------------------------------------------------------------
\subsection{Green function via general dVMC}
\label{sec:general_dVMC}

The dVMC method was first introduced in Ref.~\cite{charlebois_single-particle_2020-1} as a technique to compute the Green function from VMC for periodic clusters with translational invariance.
In order to use this approach in the context of embedding techniques (Cluster perturbation theory, Cluster dynamical mean-field theory, etc.), it is essential to relax the constraint of periodic boundary conditions and translational invariance. Here we introduce the general dVMC method that can be applied to clusters without periodic boundary conditions or translational invariance. In section~\ref{sc:translation} we highlight the differences between the general dVMC method and the original constrained version
of the technique presented in Ref.~\cite{charlebois_single-particle_2020-1}.

In order to measure the excited sectors of the Hamiltonian ($N_e+1$ and $N_e-1$), we build a collection of excited states: $\hat{c}^\dag_{i\sigma} \ket{\psi_{im\sigma}}$ for the $N_e+1$ sector and $\hat{c}_{i\sigma} \ket{\psi_{im\sigma}}$ for the $N_e-1$ sector, where we have introduced $\ket{\psi_{im\sigma}}=\hat{B}_{im\sigma}\ket{\Omega}$. The operator $\hat{B}_{im\sigma}$ can be any $N_e$ conserving operator. By experience, we have determined that the following set of operators,
\begin{eqnarray}
&&\hat{B}_{i0\sigma} = \mathbb{1} 
\label{eq:exc_states1}
\\
&&\hat{B}_{i1\sigma} = \hat{n}_{i\bar{\sigma}} 
\label{eq:exc_states2}
\\
&&\hat{B}_{im\sigma} = \hat{n}_{b_{im},\bar{\sigma}}\hat{n}_{b^\prime_{im},\sigma}, \quad\quad\quad m \geq 2,
\label{eq:exc_states}
\end{eqnarray}
is ``effective'', meaning it is not complete nor orthonormal, and can be computed efficiently within our VMC implementation. Here $i$ is the site index, $\sigma$ is the spin index and $m$ is the excitation label. The first two states in this set of excitations ($m=0$ and $m=1)$ have no degree of freedom, but for $m\geq 2$ the operator $\hat{B}_{im\sigma}$ contains parameters $b_{im}$ and $b^\prime_{im}$ that depend on $m$. This choice of $\ket{\psi_{im\sigma}}$ is flexible enough to produce a sufficiently large number of independent states for $m = 2,\hdots, N_{\rm exc}-1$, where $N_{\rm exc}$ is the number of excitations per site. The $b^{(\prime)}_{im}$ are chosen so that they are within a given number of nearest-neighbor hops from the site $i$. Typically a range of two to three hops is used, which yields $N_{\rm exc}\sim $\,40\,--100. A more detailed discussion and an example of the excitation scheme can be found in Appendix ~\ref{sec:AppendixB}. Heuristically, we expect that the excited state with one more particle at position $i$ is affected by the site occupations in the ground state only over a finite distance of $i$.

It is worth repeating that the purpose of this set of excited states is to provide a basis to expand the excitation sector of the Hamiltonian. %This is similar to the band Lanczos expansion
This is similar to the Krylov subspace used in the band Lanczos method~\cite{freund2000,aichhorn2006a}, but here the effective basis in which we expand the Hamiltonian contains some freedom to choose the order, the type, and the number, $N_{\rm exc}$, of excitations. However, by convention we take the $m=0$ state to be trivial ($\hat{B}_{i0\sigma} = \mathbb{1}$), a constraint that will prove useful later.

With this basis for excited states we can measure the excited sectors of the Hamiltonian as,
\begin{align}
H^+_{im\sigma,jn\sigma^\prime} &= \bra{\psi_{im\sigma}} \hat{c}_{i\sigma}\hat{H} \hat{c}^\dagger_{j\sigma^\prime} \ket{\psi_{jn\sigma^\prime}}, \label{eq:He_k} \\
H^-_{im\sigma,jn\sigma^\prime} &= \bra{\psi_{im\sigma}} \hat{c}^\dagger_{i\sigma}\hat{H} \hat{c}_{j\sigma^\prime} \ket{\psi_{jn\sigma^\prime}}. \label{eq:Hh_k}
\end{align}
The main difference between this approach and the band Lanczos method is that the effective basis chosen here is not orthonormal, thus
the overlap matrix is not equal to the identity matrix, an needs to be measured:
\begin{align}
S^+_{im\sigma,jn\sigma^\prime} &= \bra{\psi_{im\sigma}} \hat{c}_{i\sigma}\hat{c}^\dagger_{j\sigma^\prime} \ket{\psi_{jn\sigma^\prime}}, \label{eq:Se_k}\\ 
S^-_{im\sigma,jn\sigma^\prime} &= \bra{\psi_{im\sigma}} \hat{c}^\dagger_{i\sigma}\hat{c}_{j\sigma^\prime} \ket{\psi_{jn\sigma^\prime}}. \label{eq:Sh_k}
\end{align}
where $i=0,\hdots,N-1$, $m=0,\hdots,N_{\rm exc}-1$, and $\sigma=\pm 1$.
It is useful to express the l.h.s. of Eq.~(\ref{eq:He_k}-\ref{eq:Sh_k}) in the more compact matrix notation, $\mathbf{S}^{\pm}$ and $\mathbf{H}^{\pm}$,
where the size of the vector space is set by the combination of $i$, $m$ and $\sigma$.
%the number of lattice sites $i=0,\hdots,N-1$, the number of excitations per site $m=0,\hdots,N_{\rm exc}-1$, and $\sigma=\pm 1$. 
Hence, the dimension of $\mathbf{S}^\pm$ and $\mathbf{H}^\pm$ is $2 N N_{\rm exc}$, or only $N N_{\rm exc}$ if we focus on one spin only. 
With the exception of the Green function matrix, we omit the $\pm$ superscript in the remainder of this section in order to maintain a concise notation. 

Now that we have the overlap matrix $\mathbf{S}$ in this non-orthogonal basis, we can define an abstract Green function matrix of the same dimension as $\mathbf{S}$ and $\mathbf{H}$. Following Ref.~\cite{soriano_theory_2014}, we can express this matrix in the non-orthonormal basis as,
\begin{align}
\mathbf{G}^{\pm}(z) &= \mathbf{S}((z\pm\Omega)\mathbf{S}\mp\mathbf{H})^{-1}\mathbf{S}
\label{eq:Gpm1}
\\
&= \mathbf{S}^{1/2}((z\pm\Omega)\mathbb{1} \mp \mathbf{M})^{-1}\mathbf{S}^{1/2},
\label{eq:Gpm2}
\end{align}
where $\mathbf{M} \equiv \mathbf{S}^{-1/2}\mathbf{H} \mathbf{S}^{-1/2}$ is Hermitian. The matrix $\mathbf{M}$ can be expressed using its eigenvalue decomposition as $\mathbf{M} = \mathbf{U} \mathbf{E} \mathbf{U}^\dag$. This step helps to speed up the matrix inversion done at each complex frequency $z$. Finally, for the Green function matrix we have,
\begin{align}
\mathbf{G}^{\pm}(z) &= 
\mathbf{S}^{1/2}\mathbf{U}((z\pm\Omega)\mathbb{1} \mp\mathbf{E})^{-1}\mathbf{U}^\dag \mathbf{S}^{1/2}
\label{eq:Gpm3}
\\
&= \mathbf{Q}((z\pm\Omega)\mathbb{1} \mp \mathbf{E})^{-1}\mathbf{Q}^\dag,
\label{eq:Gpm4}
\end{align}
where $\mathbf{Q}\equiv \mathbf{S}^{1/2}\mathbf{U}$ (recall that $\mathbf{S}$ is Hermitian). 
The matrix $\mathbf{S}$ has the eigenvalue decomposition $\mathbf{S} = \mathbf{V} \mathbf{D}\mathbf{V}^\dag$, and the square root or inverse square root of $\mathbf{S}$ is obtained
by taking the square root of the elements of the diagonal matrix $\mathbf{D}$.

%-------------------------------------------------------------------------------
\subsection{Filtering algorithm}

In principle, the overlap matrix $\mathbf{S}$, as defined in Eqs.~\eqref{eq:Se_k} and~\eqref{eq:Sh_k}, is positive definite. In practice however, we obtain only Monte Carlo estimates of these matrix elements, which are not constrained to satisfy this condition. Consequently, negative eigenvalues can emerge in the spectrum of $\mathbf{S}$.
All that is required to reduce the effect of the Monte Carlo noise and restore the positive definiteness of $\mathbf{S}$ is to truncate the matrix $ \mathbf{D}$ (the spectrum of $\mathbf{S}$) and keep only the positive eigenvalues, resulting in a smaller matrix $\mathbf{\bar D}$. %In order to obtain a similar matrix to $\mathbf{S}$, but filtered, 
We do a similar truncation for the columns of $\mathbf{V}$, resulting in a rectangular matrix $\mathbf{\bar V}$. With these truncated matrices we can construct the filtered overlap matrix $\mathbf{\bar S} = \mathbf{\bar V} \mathbf{\bar D} \mathbf{\bar V}^\dag$.

This filtering is done before the operations in Eqs.~\eqref{eq:Gpm1}-\eqref{eq:Gpm4}: we replace only $\mathbf{S}$ by its filtered version $\mathbf{\bar S}$, but it affects these four equations and the resulting matrices $\mathbf{M}$, $\mathbf{U}$, $\mathbf{E}$ and $\mathbf{Q}$.

Note that we refer to the matrix $\mathbf{\bar S}$ as {\it filtered} and not {\it truncated} because it has the same dimension as the original matrix $\mathbf{S}$. For the sake of speed, one last simple optimization can be done on the spectrum $\mathbf{\bar E}$ itself. Indeed, truncating $n_c$ dimensions out of the matrix $\mathbf{D}$ results in $n_c$ null vectors in $\mathbf{Q}$ and $n_c$ null eigenvalues in $\mathbf{E}$ that can be removed, without affecting the result.

%-------------------------------------------------------------------------------
\subsection{Green function}

Let us recall that the matrix $\mathbf{G}^{\pm}(z)$ is an abstract construction and to obtain the Green function requires one last step. The indices of our matrix notation are $im\sigma$, corresponding to site, excitation, and spin, but the Green function that we are looking for does not contain any excitation index $m$. Since we chose the first element $m=0$ of the vector in Eq.~\eqref{eq:exc_states} to be the trivial excitation, the Green function is obtained by summing the electron and hole Green functions and keeping only $m=n=0$:
\begin{align}
G_{ij,\sigma}(z) = \left[ \mathbf{G^+}(z) + \mathbf{G^-}(z) \right]_{ij,\sigma=\sigma^\prime,m=n=0}.
\end{align}
This treatment yields a Green function equivalent to summing the hole and electron Green function in the Lehmann representation:
\begin{align}
G^-_{ij,\sigma}(z) &= \sum_\ell \frac{\bra{\Omega}\hat{c}^\dagger_{i\sigma}\ket{E^-_{\ell}}\bra{E^-_{\ell}} \hat{c}_{j\sigma} \ket{\Omega}} {z - \Omega + E^-_{\ell}}, \label{eq:Gh} \\
G^+_{ij,\sigma}(z) &= \sum_\ell \frac{\bra{\Omega}\hat{c}_{i\sigma}\ket{E^+_{\ell}}\bra{E^+_{\ell}} \hat{c}^\dagger_{j\sigma} \ket{\Omega}} {z + \Omega - E^+_{\ell}}, \label{eq:Ge}
\end{align}
where the states $\ket{E^\pm_{\ell}}$ are obtained by solving the generalized eigenvalue problem $\mathbf{H}\ket{E_{\ell}} = E_{\ell}\mathbf{S}\ket{E_{\ell}}$. However, in many cases, as in VMC, the ground state is not known explicitly, so we can only know $\ket{E^\pm_{\ell}}$ via its projection onto other states. This is the case here, where in Eq.~\eqref{eq:Gpm4} the rectangular matrices $Q^\pm_{im\sigma,\ell}$, when evaluated at $m=0$, represent the projection $\bra{\Omega}\hat{c}^{(\dag)}_{i\sigma}\ket{E^\pm_{\ell}}$. When we evaluate the Green function on the real axis $\omega$, $z=\omega+i\eta$ where $\eta$ is a small Lorentzian broadening.

%-------------------------------------------------------------------------------
\subsection{Differences with translation invariant dVMC}
\label{sc:translation}
In the translationally invariant case, the matrices introduced in this section ($\mathbf{G}$, $\mathbf{S}$ and $\mathbf{H}$) can be Fourier transformed and become block diagonal (diagonal in $\mathbf k$). Each $\mathbf k$ point can then be treated separately, and the matrices that need to be sampled with Monte Carlo and diagonalized are only of size $2N_{\rm exc} \times 2N_{\rm exc}$. 

Additionally, we have introduced here a new noise filtering approach, which is essential to the general dVMC algorithm presented here, and can likely improve results for translationally invariant systems with periodic boundary conditions, though it did not prove to be important in the previous study~\cite{charlebois_single-particle_2020-1}.

%-------------------------------------------------------------------------------
\subsection{Cluster perturbation theory (CPT)}
\label{sc:CPT}

Cluster perturbation theory is a technique designed to compute spectral properties of
strongly-correlated systems \cite{gros_cluster_1993, senechal_spectral_2000,Senechal2002}.
The central idea behind the approach is to construct a superlattice of
clusters that are coupled by intercluster hopping terms. The cluster problem
is solved with dVMC as the impurity solver to obtain the cluster self energy, from which the CPT 
Green function can be obtained according to,
\begin{equation}
G(\mathbf{k},\omega) = \frac{1}{N}\sum_{ij}e^{-i\mathbf{k}\cdot(\mathbf{R}_i-\mathbf{R}_j)}
\left[\frac{1}{\omega+i\eta-\mathbf{t}(\mathbf{k})-\boldsymbol{\Sigma}_c(\omega)}\right]_{\mathbf{R}_i\mathbf{R}_j},
%G_{\mathbf{R},\mathbf{R}^\prime}(\tilde{\mathbf{k}},\omega).
\end{equation}
where $\boldsymbol{\Sigma}_c(\omega)$ is the cluster self-energy and $\mathbf{t}(\mathbf{k})$ is the hopping matrix, with
elements,
$t_{ij}(\mathbf{k})=\sum_{\mathbf{a}}e^{-i\mathbf{k}\cdot\mathbf{a}}t_{\mathbf{a}+\mathbf{R}_i,\mathbf{R}_j}$,
where the sum is over all superlattice vectors, $\mathbf{a}$, and $\mathbf{R}_i,\mathbf{R}_j$ are the positions of the sites within each cluster. Note that the matrix dimension of $\boldsymbol{\Sigma}_c(\omega)$ and $\mathbf{t}(\mathbf{k})$ differs from the matrix dimensions of the previous sections.
This approach yields a periodized and translationally invariant Green function with arbitrary $\mathbf{k}$-point resolution.

This procedure can be seen as reconstructing the full lattice by tiling the cluster's self-energy. Although the non-interacting part of the Hamiltonian $t_{ij}(\mathbf{k})$ is exact, the self-energy $\Sigma_{c,ij}(\omega)$ is null between clusters. In order to reduce the effect of this approximation, it is necessary to treat larger clusters. 
In the results section to follow, we study the pseudogap phase of the Hubbard model for large clusters, in order to investigate the existence and nature of the Fermi arcs. Benchmarking is provided in the appendices. 

%===============================================================================
\section{Results}
\label{se::results}

We first introduce the model and then study in turn the doping dependence, cluster size and then shape dependence and finally discuss the computational cost. 

%-------------------------------------------------------------------------------
\subsection{Model}
We apply this new technique to the Hubbard model:
\begin{equation}
\hat{H} = \sum_{ij,\sigma} (t_{ij} \hat{c}^\dagger_{i\sigma}\hat{c}_{j\sigma} + \text{h.c.}) + U\sum_i \hat{n}_{i\uparrow} \hat{n}_{i\downarrow},
\end{equation}
where $t_{ij}$ includes first ($t$), second ($t^\prime$) and third ($t^{\prime\prime}$) nearest neighbor hoppings. In the results
that follow we choose $(t,t^\prime,t^{\prime\prime})=(-1.0,0.3,-0.2)$, which is a parameter set believed to be relevant the case of
certain cuprate superconductors \cite{Andersen1995,senechal_hot_2004-1}. 
We treat clusters of different sizes, shapes and doping, up to $N=64$ sites and $N_e=64$ electrons.
Unless otherwise indicated, we set the Hubbard interaction strength to $U=8$. We used a Lorentzian broadening factor $\eta=0.1$.

In the following three sections we present a systematic study of the spectral function
of the 2D Hubbard model computed using CPT 
\cite{gros_cluster_1993, senechal_spectral_2000, Senechal2002} with dVMC as the impurity solver. 
We have also performed a set of benchmarks on small 
clusters that are within reach of exact diagonalization solvers in order to calibrate the accuracy of the general
dVMC method. These benchmarks are presented in Appendix~\ref{sec:AppendixA}.

%-------------------------------------------------------------------------------
\subsection{Doping dependence}

We begin by investigating the behavior of the Fermi surface as a function of hole
doping. We treat square clusters of three different sizes, $N=16,36, 64$.

\begin{figure*}
    \begin{center}
           \includegraphics[width=0.98\textwidth,trim={1cm 2.5cm 1cm 1cm},clip]{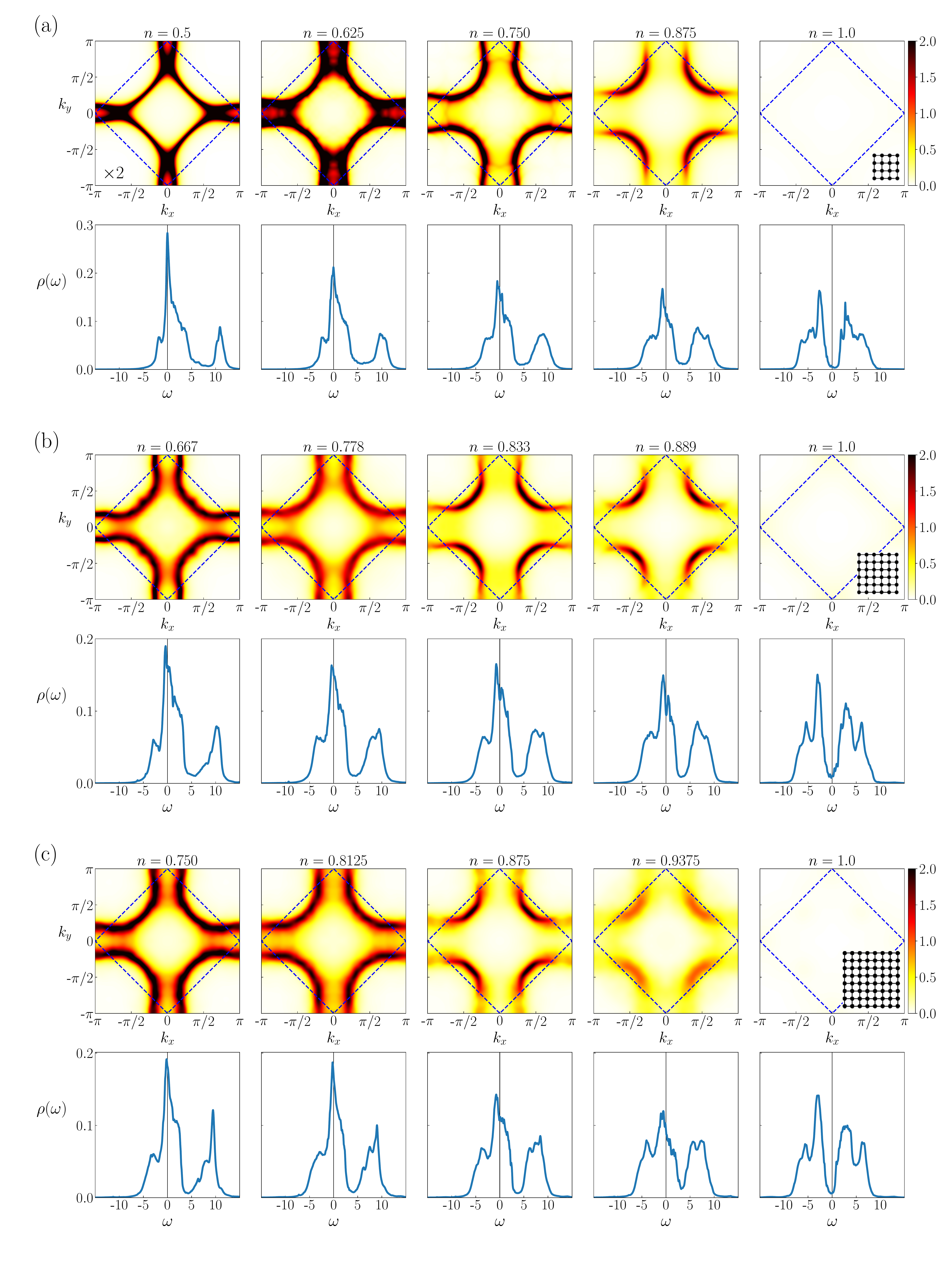}
    \end{center}
    \caption{Fermi surface (top row) and density of states $\rho(\omega)$ (bottom row) for different dopings $n$ for a (a) 4$\times$4, (b) 6$\times$6 and (c) 8$\times$8 cluster. The blue dashed line shows the antiferromagnetic zone boundary.}
    \label{fig:FS_vs_doping}
\end{figure*}

Beginning with the smallest cluster [Fig.~\ref{fig:FS_vs_doping}(a)], at large doping we observe a roughly
cylindrical Fermi surface consistent with the system being in the normal metallic phase. As the density is increased (the number of holes is reduced) the Fermi surface
loses intensity and the spectral weight becomes more concentrated in the nodal region. 
At a density of $n=0.875$ 
there is a loss of spectral weight in the anti-nodal regions and clear evidence of the formation of Fermi arcs, 
suggesting that the system has entered the pseudo-gap phase. At half-filling ($n=1.0$) the Fermi level lies in the Mott gap 
and there is no significant spectral weight at the Fermi surface.

This transition is also evident in the density of states. 
At half-filling the Mott gap is opened, but as the density is decreased (hole doping increased) the gap closes and
a Van Hove singularity forms at the Fermi level. We find a similar evolution for the cluster of 36 sites  [Fig.~\ref{fig:FS_vs_doping}(b)],  
as well as the cluster of 64 sites  [Fig.~\ref{fig:FS_vs_doping}(c)].
The important qualitative features of both the Fermi
surface and the density of states are in good agreement for each cluster size, with only some quantitative differences evident. 

%...............................................................................
%...............................................................................

%-------------------------------------------------------------------------------
\subsection{Cluster size dependence}

We proceed with a comparison of the Fermi surface for clusters of two different
sizes at the same density, in order to examine the dependence on cluster size. 
We compare results for a cluster of 16 sites [Figs. \ref{fig:FS_vs_doping}(a)] 
to a cluster of 64 sites [Figs. \ref{fig:FS_vs_doping}(c)] respectively.

At a density of $n=0.750$ we observe a cylindrical Fermi surface for both cluster sizes, with 
a region of slightly higher spectral weight appearing to develop in the nodal region.
The concentration of spectral weight in the nodal region becomes more evident 
at a density of $n=0.875$, for both clusters, though it is somewhat more concentrated
near the node for the 64 site cluster. 
At half-filling both clusters show negligible spectral weight at the Fermi level, consistent with the density of states, 
which shows that the Fermi level indeed lies the gap.
Generally, we observe that the behavior of the spectral function and the density of
states is not strongly dependent on system size, larger clusters retain the same
qualitative features as smaller clusters with some quantitative differences.
However, the cluster geometry does appear to play a more important
role, as we discuss in the following section.

%-------------------------------------------------------------------------------
\subsection{Cluster shape and size dependence}

\begin{figure*}[!ht]
    \begin{center}
           \includegraphics[width=0.99\textwidth,trim={0.2cm 0.15cm 0.1cm 0.2cm},clip]{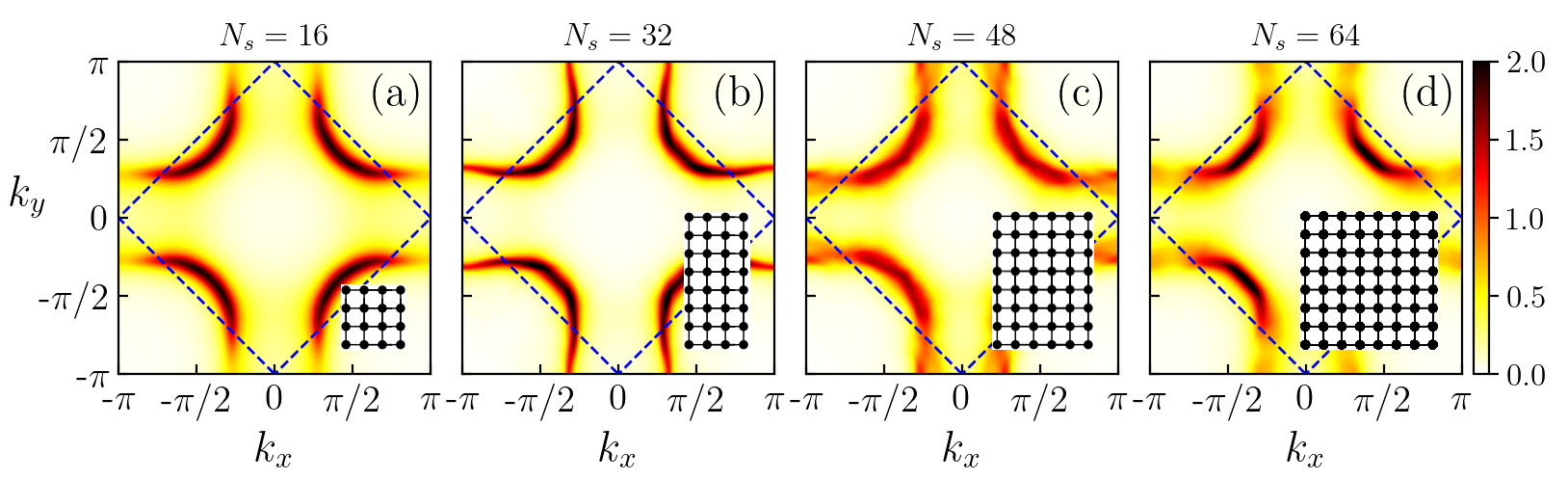}
    \end{center}
    \caption{Fermi surface versus cluster size and shape for a fixed density $n=0.875$. We note that all of the results have been symmetrized.}
     \label{fig:FS_vs_L_n0.875}
\end{figure*}

\begin{figure*}[!ht]
    \begin{center}
           \includegraphics[width=0.99\textwidth,trim={0.2cm 0.15cm 0.1cm 0.2cm},clip]{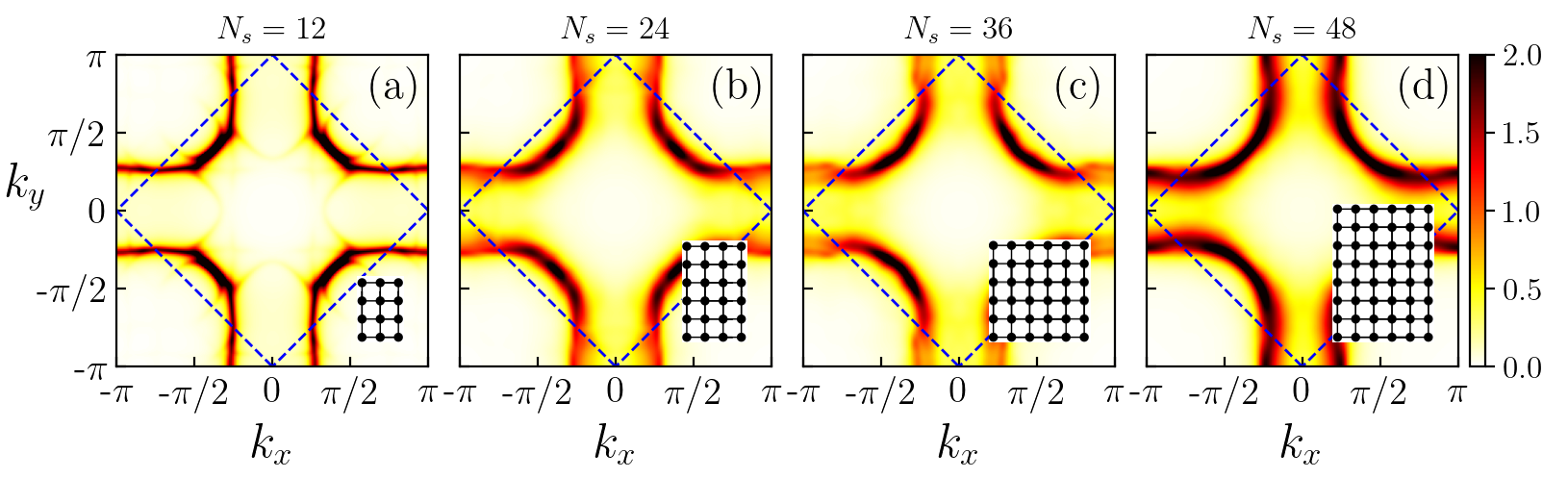}
    \end{center}
    \caption{Fermi surface versus cluster size and shape for a fixed density $n=0.8333$. All results have been symmetrized as in Fig.~\ref{fig:FS_vs_L_n0.875}.}
    \label{fig:FS_vs_L_n0.833}
\end{figure*}

Having examined the behavior of the Fermi surface as a function
of doping as well as cluster size we now consider the effect of
cluster geometry. We symmetrize the Fermi sufaces by 
averaging the reflection with respect to diagonals $k_x = \pm k_y$.
In Fig.~\ref{fig:FS_vs_L_n0.875} we show the spectral weight at the Fermi level
for different clusters with $n=0.875$. We observe
that the rectangular clusters have more spectral weight near the edges of the BZ
than the square clusters, whose Fermi surfaces are more consistent with Fermi arcs.
A similar effect is visible at $n=0.8333$, where the rectangular clusters 
[Fig.~\ref{fig:FS_vs_L_n0.833}(a),(b),(d)] have relatively
more spectral weight near the edges of the BZ than the square cluster
[Fig.~\ref{fig:FS_vs_L_n0.833}(c)], which shows clearer evidence of the formation
of Fermi arcs. This effect is also observed in the 12 site ED calculation. In fact, our dVMC result 
[Fig.~\ref{fig:FS_vs_L_n0.833}(a)] reproduces exactly the result shown in Fig.~3 (at $U=8$) of Ref.~\cite{senechal_hot_2004-1}. 

%...............................................................................
%...............................................................................
%...............................................................................

%...............................................................................

%-------------------------------------------------------------------------------
\subsection{Computational cost}

In this section, we provide an estimate of the typical computational cost
for two system sizes, the first a 4$\times$4 cluster (ED and dVMC) and the second an 8$\times$8 cluster (dVMC only). Calculations were performed on an AMD 7532 CPU (2.40 GHz).

For the 4$\times$4 system [Fig.\ref{fig:FS_vs_doping}(a) $n=1.0$], the dVMC calculation of the ground state and the excitations required 3.5 hours for each step. Sampling was done with 10 MPI tasks on 8 OpenMP processes. Each MPI task required 0.5 Go. In comparison, the full ED calculation takes around 30.5 hours on a single core, and requires above 500 Go in memory (making it hard to parallelize). For this example, though the computation time is similar, the memory needed for dVMC is much smaller than ED.

Because of the exponential scaling in ED, it is not possible to treat systems much larger than 16 sites. But the scaling is considerably better for dVMC (polynomial in cluster size and variational parameters) and we can reach 64 sites. The result in Fig.\ref{fig:FS_vs_doping}(c) ($n=1.0$) required 45 and 20 hours of computation, respectively, for the ground state and excitations. Sampling was done with 10 MPI tasks on 10 OpenMP processes, with each MPI task requiring 4 Go of memory.

%===============================================================================
\section{Discussion}
\label{se::discussion}

The technique has several important advantages. First, it has favorable scaling with
system size, which enables the treatment of systems beyond the capacity of current
exact diagonalization solvers, typically limited to about $16$ sites near half-filling.
Second, the method is quite flexible and can be straightforwardly adapted to 
make use of any variational wavefunction; it is not constrained to the Slater-Jastrow-Gutzwiller ground state used here. 
For instance, one novel choice would be artificial Neural Network ansatzes inspired by ideas from machine learning \cite{Han2019, Choo2020,Pfau2020,Stokes2020}.

The key idea here is that we never express any state in the full $4^{N}$ Hilbert space. It is only necessary to calculate a set of projections
($\langle\Omega|\hat H| x\rangle$ and ${\langle x | \Omega\rangle}$) on the variational ground state $\ket{\Omega}$
in order to obtain a good approximation of the true ground state and 
the excitation spectrum (Eq.~\eqref{eq:He_k}-\eqref{eq:Sh_k}). This is in contrast to the traditional Lanczos and band Lanczos methods 
where a few vectors of the full Hilbert space need to be computed before obtaining the low dimension effective Hamiltonian, thus limiting studies 
to cluster of about $20$ or even fewer sites. This constraint is removed with the present Green function calculation algorithm. It can be used with any 
ground state from which we can calculate static observables.

As a demonstration of the capabilities of the technique, we have performed an extensive set
of CPT calculations on the 2D Hubbard model. 
Our results show good agreement with exact diagonalization impurity solvers for small clusters, and
for large clusters they capture the important features of the Fermi surface as a function of hole doping, such as 
the formation of Fermi arcs marking the transition to the pseudogap, a result that has been disputed for small
clusters~\cite{verret_fermi_2022-1}. We treat large clusters of up to 64 sites, beyond the reach of current ED impurity solvers, to examine 
the dependence on cluster size, and find that the results are converged for clusters of 16 sites or more, with
only minor quantitative differences evident between the Fermi surfaces at $N=16$ and $N=64$. 

We do however observe some dependence on the shape of the cluster, with square clusters giving the clearest
indication of the formation of Fermi arcs. Although this effect is consistent with ED results \cite{senechal_hot_2004-1}, it is clearly exposed here 
by our dVMC results for clusters of various shapes and sizes. This effect is likely caused either by the asymmetry of the superlattice or by the
symmetrization of the Fermi surface that we impose. This suggests that it is preferable to use square clusters in order to study the pseudogap transition in 
future quantum cluster method calculations.

Our observation of Fermi arcs for various system sizes provides convincing evidence of their existence in
the doped Hubbard model, a notable and important result considering that the origin and robustness of these features remain
an open question, as noted in Refs.~\cite{verret_fermi_2022-1,proust2019}. Reference~\cite{verret_fermi_2022-1} demonstrated that it is not possible to distinguish between the formation of hole-pockets and Fermi arcs for small clusters.
Though we have used the G-scheme periodization in this work, the tendency towards the formation of Fermi arcs for very large clusters 
offers compelling evidence that the existence of these features is not strongly dependent on the periodization scheme.
As discussed in Sec.~\ref{sc:CPT}, the effect of neglecting the self-energy between clusters, as is done in CPT, diminishes with increasing cluster size.

Fig.~\ref{fig:FS_vs_doping}(c) reproduces precisely what is observed in ARPES measurements (e.g. Fig.~20 of Ref.\cite{sobota_angle-resolved_2021}). In other words, in the paramagnetic Hubbard model we observe a Mott insulator at half-filling, Fermi arcs in the underdoped region and large Fermi surface in th overdoped region. We observe only Fermi arcs and no trace of hole-pockets. 
This confirms that the single-layer, one-band Hubbard model, in the paramagnetic state, contains the minimal ingredients to produce the Fermi arcs as observed in ARPES.

The existence of these Fermi arcs is in direct contradiction with the hole-pocket picture, suggested by quantum oscillations experiments~\cite{doiron-leyraud_quantum_2007, bangura_small_2008,leboeuf_electron_2007,yelland_quantum_2008,jaudet_haas--van_2008,kunisada_observation_2020}. Hole pockets are  expected in a metallic antiferromagnetic state close to half-filling, a case we did not consider. Similarly, oscillations can come from pockets that appear in other broken symmetry states, such as charge order~\cite{Sebastian_Harrison_Liang_Bonn_Hardy_Mielke_Lonzarich_2012}. Fermi arcs should not produce any quantum oscillation signal due to the incoherent nature of the antinodal region, as seen in Figs.~\ref{fig:FS_vs_doping}, \ref{fig:FS_vs_L_n0.875} and \ref{fig:FS_vs_L_n0.833}.  This is confirmed by the experiments of Ref.~\cite{kunisada_observation_2020} that observed both arcs and pockets in the same 5-layer material with both ARPES and quantum oscillations but where quantum oscillations are explained by the observed hole-pockets, with no Fermi arc contribution. Indeed, quantum oscillations require coherent transport of electrons around the Fermi surface. Most likely no electron accomplish even a single period around the Fermi surface before getting scattered in the antinodal region, resulting in the absence of a quantum oscillation signal from Fermi arcs.
 
%===============================================================================
\section{Conclusion}

In this work we have introduced and benchmarked an extension
of the original dVMC method, which is designed to compute the Green function for strongly-correlated systems. This new method removes the constraint of periodic boundary conditions and translational invariance that was a requirement of the original dVMC technique \cite{charlebois_single-particle_2020-1}. In addition, we have introduced a filtering algorithm to reduce Monte Carlo noise and improve the quality of the results. The difference between exact diagonalization results and our approach is of order 5\% at most for the largest clusters available.

The method is free of the sign problem
and does not require analytic continuation. In principle, there is no restriction on the choice of strongly correlated Hamiltonian, and it can be used to 
study any magnetic phase with equivalent precision for large and small interaction. It can be implemented with complex numbers, for instance to treat spin-orbit coupling. 
One additional extension reserved for future work is to adapt the formalism to measure the Nambu Green function, in order to treat systems with superconducting phases. The formalism is sufficiently versatile 
to be used to measure the Green function for any kind of variational ground state, including Neural Network ansatzes.
Finally, the features discussed here make the approach applicable to a variety of embedding techniques, including cluster dynamical mean-field theory (CDMFT), and dynamical cluster approximation (DCA), among others.

By applying this generalized dVMC approach with Cluster perturbation theory on unprecedently large cluster sizes, we claim that we have demonstrated that Fermi arcs with no trace of hole-pockets can form near half-filling in the single-band Hubbard model.

\begin{acknowledgments}

We thank Nicolas Gauthier and Louis Taillefer for useful discussions.
This research was undertaken thanks in part to funding from the Canada First Research Excellence Fund, the Natural Sciences and Engineering Research Council (Canada) under Grant No. RGPIN-2021-04043 and 
RGPIN-2019-05312. P.R. was supported by a postdoctoral fellowship from Institut quantique”.
Computing resources were provided by Compute Canada and Calcul Qu\'ebec. 

\end{acknowledgments}

\appendix
%===============================================================================
\section{Benchmarks against exact diagonalization}
\label{sec:AppendixA}

\begin{figure}[!ht]
    \begin{center}
           \includegraphics[width=0.8\columnwidth]{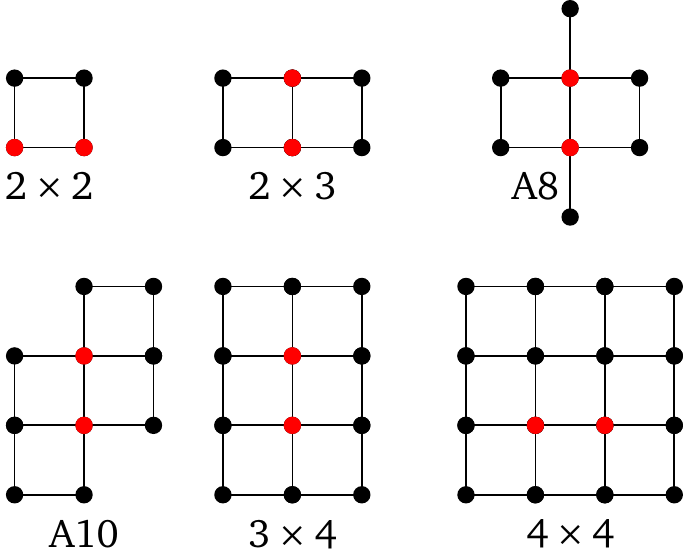}
    \end{center}
    \caption{Cluster geometries. For each cluster, the pair of sites drawn in red indicates the
    bond along which the spin-spin correlation is computed in Table \ref{table:ED_vs_VMC_high_stats}. \label{fig:clusters}
}
\end{figure}

To gauge the accuracy of the technique we perform a set of benchmarks
against exact diagonalization on small clusters of various sizes and shapes.
We begin with a comparison
of the VMC and ED ground states for clusters up to $N=16$ sites (Fig.~\ref{fig:clusters})
at and away from half-filling. In table \ref{table:ED_vs_VMC_high_stats} we
present results for the ground state energy, double occupancy and spin-spin
correlation functions for both techniques for all the clusters in  Fig.~\ref{fig:clusters}. At half-filling the VMC ground state
energy shows excellent agreement with the ED result, generally to within 1\%,
while the double occupancy and spin-spin correlation function are accurate to
within 5\%. Away from half-filling the ground state energy is accurate to within
5\% and the double occupancy and spin-spin correlation is generally within 15\%
of the ED result.

%...............................................................................
%...............................................................................

%...............................................................................
\begin{table*}[ht!]
	\begin{tabular}{ll | llllllllllllllllll}
		\hline\hline
		\multicolumn{1}{c}{} && \multicolumn{1}{c}{$N_e$} &&  \multicolumn{1}{c}{$E_\textmd{ED}/N$} && \multicolumn{1}{c}{$E_\textmd{VMC}/N$}  && \multicolumn{1}{c}{$D_\textmd{ED}$} && \multicolumn{1}{c}{$D_\textmd{VMC}$}  && \multicolumn{1}{c}{$\langle S^z_i S^z_j\rangle_\textmd{ED}$} && \multicolumn{1}{c}{$\langle S^z_i S^z_j\rangle_\textmd{VMC}$}  \\ \hline
		2$\times$2 && 4 && -4.3300587 && -4.3300576 &&  0.032496  && 0.03268(38) && -0.6128  &&  -0.6127(14) \\
		2$\times$3 && 6 && -4.36297 &&  -4.36168(39) && 0.042213 && 0.04194(17) && -0.42790 && -0.43290(65) \\
		A8 &&   8 && -4.34301 && -4.34120(10) && 0.054275 && 0.05426(98) && -0.2277 && -0.2313(24)\\
		B10 &&  10 &&-4.385580 && -4.381509(72) && 0.052058 && 0.05305(80) && -0.3476 && -0.3407(19)\\
		3$\times$4 && 12 &&-4.40944 &&  -4.403467(59) &&  0.050780  && 0.05112(86) && -0.3925 && -0.4056(23)\\
		4$\times$4 && 16 && -4.42553 && -4.419112(52) && 0.051814 && 0.0530(11) && -0.372674 && -0.3892(26) \\
		\hline
		2$\times$2 && 2 && -0.8019377  &&  -0.8019377  &&  0.008648  &&  0.008549(63)  &&  -0.13578  &&  -0.13523(26)  &&  \\
		2$\times$3 && 4 &&  -0.76348  &&  -0.763390(12) &&  0.01436  &&  0.01427(12)  &&  -0.19466  &&  -0.19986(45)  &&  \\
		A8 &&  6 && -0.67643  &&  -0.671260(60) &&  0.02798  &&  0.02941(19)  &&  -0.07254  &&  -0.07318(49)  &&  \\  
		B10 &&  8 &&  -0.6824  &&  -0.669543(78) &&  0.02943  &&  0.03095(20)  &&  -0.03942  &&  -0.05782(50)  &&  \\
		3$\times$4 && 10 && -0.6690  &&  -0.646762(83) &&  0.02872  &&  0.03305(21)  &&  -0.05137  &&  -0.05828(56)  &&  \\
		4$\times$4 && 14 && -0.6326 && -0.60704(13) && 0.03929 && 0.04312(57) &&  -0.2478 &&  -0.2120(16) &&  \\
		\hline\hline
	\end{tabular}
	\caption{Comparison of ED and VMC ground state energy per site ($E_\textmd{ED}/N$ and $E_\textmd{VMC}/N$ respectively), double occupancy ($D_\textmd{ED}$ and $D_\textmd{VMC}$) and the $S_z$-$S_z$ correlation ($\langle S^z_i S^z_j\rangle_\textmd{ED}$ and $\langle S^z_i S^z_j\rangle_\textmd{VMC}$). The sites for the $S_z$-$S_z$ correlation
	are indicated in red in Fig.~\ref{fig:clusters}. The error estimate on the energy in the $2\times2$ case was omitted since it is very small and unreliable.}
	\label{table:ED_vs_VMC_high_stats}
\end{table*}
%...............................................................................

In table \ref{table:ED_vs_VMC_vs_U} we present a comparison of the VMC and ED ground states as a function of
interaction strength for two 16 site clusters, a $16\times1$ cluster and a $4\times4$ cluster, both at half-filling. The VMC 
ground state is generally highly accurate, with the ground state energy within 1\% of the exact result and the
observables within 5\%.

%...............................................................................
\begin{table*}[ht!]
	\begin{tabular}{ll | llllllllllllllllll}
		\hline\hline
		\multicolumn{1}{c}{}  && \multicolumn{1}{c}{$U$} && \multicolumn{1}{c}{$E_\textmd{ED}/N$} && \multicolumn{1}{c}{$E_\textmd{VMC}/N$}  && \multicolumn{1}{c}{$D_\textmd{ED}$} && \multicolumn{1}{c}{$D_\textmd{VMC}$}  && \multicolumn{1}{c}{$\langle S^z_i S^z_j\rangle_\textmd{ED}$} && \multicolumn{1}{c}{$\langle S^z_i S^z_j\rangle_\textmd{VMC}$}  \\ \hline
	\multirow{4}{*}{16$\times$1}   && 0 && -1.22974  &&  -1.22974  &&  0.2500  &&  0.2525(19)  &&  -0.1675  &&  -0.1677(14)  &&  \\
	                                                && 1 && -1.49962  &&  -1.49927(12)  &&  0.2139  &&  0.2157(21)  &&  -0.1984  &&  -0.1981(13)  &&  \\
						      && 4 && -2.55117  &&  -2.54990(27)  &&  0.1004  &&  0.0994(15)  &&  -0.3262  &&  -0.3159(26)  &&  \\
						      && 8 && -4.31482  &&  -4.31451(17)  &&  0.0367  &&  0.0360(12)  &&  -0.4168  &&  -0.4096(32)  &&  \\
		\hline
	\multirow{4}{*}{4$\times$4}   && 0 && -1.36803  &&  -1.36803  &&  0.2499  &&  0.2507(19)  &&  -0.2094  &&  -0.2097(11)  &&  \\
	                                             && 1 && -1.65222  &&  -1.65159(16)  &&  0.1962  &&  0.1964(14)  &&  -0.1873  &&  -0.1854(11)  &&  \\
						   && 4 && -2.70288  &&  -2.69918(35)  &&  0.1148  &&  0.1148(12)  &&  -0.2999  &&  -0.2915(21)  &&  \\
						   && 8 && -4.42553  &&  -4.41896(76)  &&  0.0518  &&  0.0530(11)  &&  -0.3727  &&  -0.3892(26)  &&  \\
		\hline\hline
	\end{tabular}
	\caption{Comparison of exact diagonalization and VMC ground state energy, double occupancy and $S_z$-$S_z$ correlation versus $U$. We show results for a 16$\times$1 cluster and a 4$\times$4 cluster. Notation as in Table~\ref{table:ED_vs_VMC_high_stats}. The error estimate on the energy in the $U=0$ case was omitted since it is very small and unreliable.}
	\label{table:ED_vs_VMC_vs_U}
\end{table*}
%...............................................................................

\begin{figure*}[!ht]
    \begin{center}
           \includegraphics[width=\textwidth]{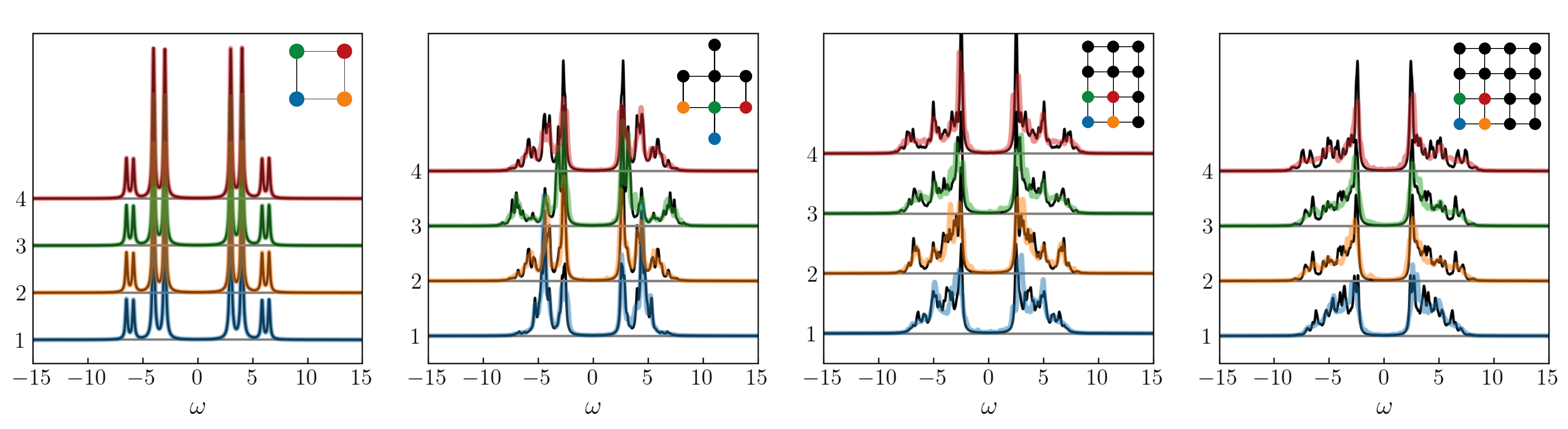}
    \end{center}
    \caption{ Comparison of the local spectral function $A_{ii}(\omega)$ computed using dVMC versus ED. The ED results are shown in black and the dVMC results in color.
    The cluster shapes are shown in the inset, with the colors denoting the sites for which the spectral function is shown in the main plot. 
     \label{fig:spectra_vs_ED_Ns_4_12_16}}
\end{figure*}

Having calibrated the accuracy of the VMC ground state we proceed with a comparison of the 
spectral function obtained using dVMC to that obtained using ED. We present several
example calculations in Fig.~\ref{fig:spectra_vs_ED_Ns_4_12_16}. For a cluster of 4 sites the
VMC result is exceptionally accurate, nearly indistinguishable from the exact result. With increasing
cluster size the VMC result continues to capture the prominent features, including the gap, with
reasonable accuracy, though it does not reproduce many of the smaller structures present in the exact solution.
We provide a numerical comparison of the spectra versus the number of excitations in Table ~\ref{table:dVMC_vs_Nexc},
which shows the results of Kolmogorov-Smirnov tests~\cite{press2007numerical} comparing the dVMC and ED spectra. 

%...............................................................................
%...............................................................................

%...............................................................................
\begin{table}[ht!]
		\begin{tabular}{ll | llllllllllll}
		\hline\hline
		\multicolumn{1}{c}{} && \multicolumn{1}{c}{$N_e$} && \multicolumn{1}{c}{$n_h = 1$} && \multicolumn{1}{c}{$n_h = 2$}  && \multicolumn{1}{c}{$n_h = 3$} && \multicolumn{1}{c}{$n_h = 4$}  \\ \hline 
		2$\times$2 && 4 && 0.00719  && 0.00690 && \multicolumn{1}{c}{-} && \multicolumn{1}{c}{-} && \\
		2$\times$3 && 6 && 0.07210 && 0.03054 && 0.02469 && \multicolumn{1}{c}{-} &&  \\
		A8 && 8 && 0.10592 && 0.09223 && 0.06395 &&  \multicolumn{1}{c}{-}  &&  \\
		B10 && 10 &&0.12057 && 0.10797 && 0.10158 && 0.08518 && \\
		3$\times$4 && 12 && 0.10141 && 0.07766 && 0.05224 && 0.04967 && \\
		4$\times$4 && 16 && 0.10933  &&  0.08581 && 0.07762  &&  0.07156   \\
		\hline
		2$\times$2 && 2  && 0.00516 && 0.00486 &&  \multicolumn{1}{c}{-} &&  \multicolumn{1}{c}{-}  && \\
		2$\times$3 && 4 && 0.06217 && 0.04108 && 0.03306 &&  \multicolumn{1}{c}{-}  &&  \\
		A8 && 6 && 0.09602 && 0.05887 && 0.04654 &&  \multicolumn{1}{c}{-} &&  \\
		B10 && 8 && 0.09608 && 0.06599&& 0.05793 && 0.05875 && \\
		3$\times$4 && 10 && 0.12825 && 0.09039 && 0.08496 && 0.08357 && \\
		4$\times$4 && 14 && 0.17240  && 0.12759 && 0.11380 && 0.10958 && \\
		\hline\hline
	\end{tabular}
	\caption{Kolmogorov–Smirnov error in the local spectral function per site versus range of excitations for different cluster geometries (see Fig.~\ref{fig:clusters}). $N_e$ is the number of electron and  $n_h$ is the number of nearest-neighbor hops allowed in the selection of the excitation operator $\hat{B}_{im\sigma}$. See Eq.~\eqref{eq:exc_states} and Appendix~
\ref{sec:AppendixB}. }
	\label{table:dVMC_vs_Nexc}
\end{table}
%...............................................................................

%...............................................................................
\begin{table}[ht!]
		\begin{tabular}{ll | llllllllllll}
		\hline\hline
		\multicolumn{1}{c}{} && \multicolumn{1}{c}{$U$} && \multicolumn{1}{c}{$n_h = 1$} && \multicolumn{1}{c}{$n_h = 2$}  && \multicolumn{1}{c}{$n_h = 3$} && \multicolumn{1}{c}{$n_h = 4$}  \\ \hline 
		\multirow{3}{*}{16$\times$1} && 1 && 0.03159  && 0.02633  && 0.02767  &&  0.02840 \\
						   && 4 && 0.05730  && 0.05176 && 0.05102 &&  0.05089  \\
						   && 8 &&  0.05821 &&  0.04807 && 0.04485   && 0.04412 \\
		\hline
		\multirow{3}{*}{4$\times$4} && 1 && 0.03435 &&  0.03218  &&  0.03296  && 0.03501 \\
						   && 4 && 0.08105 &&  0.06304  &&  0.06372  &&  0.06311 \\
						   && 8 && 0.10933  &&  0.08581 && 0.07762  &&  0.07156   \\
		\hline\hline
	\end{tabular}
	\caption{Kolmogorov–Smirnov error  in the local spectral function per site versus $U$. Notation as in Table~\ref{table:dVMC_vs_Nexc}.}
	\label{table:dVMC_vs_Nexc_vs_U}
\end{table}
%...............................................................................

\begin{figure}[!ht]
    \begin{center}
           \includegraphics[width=1.\columnwidth]{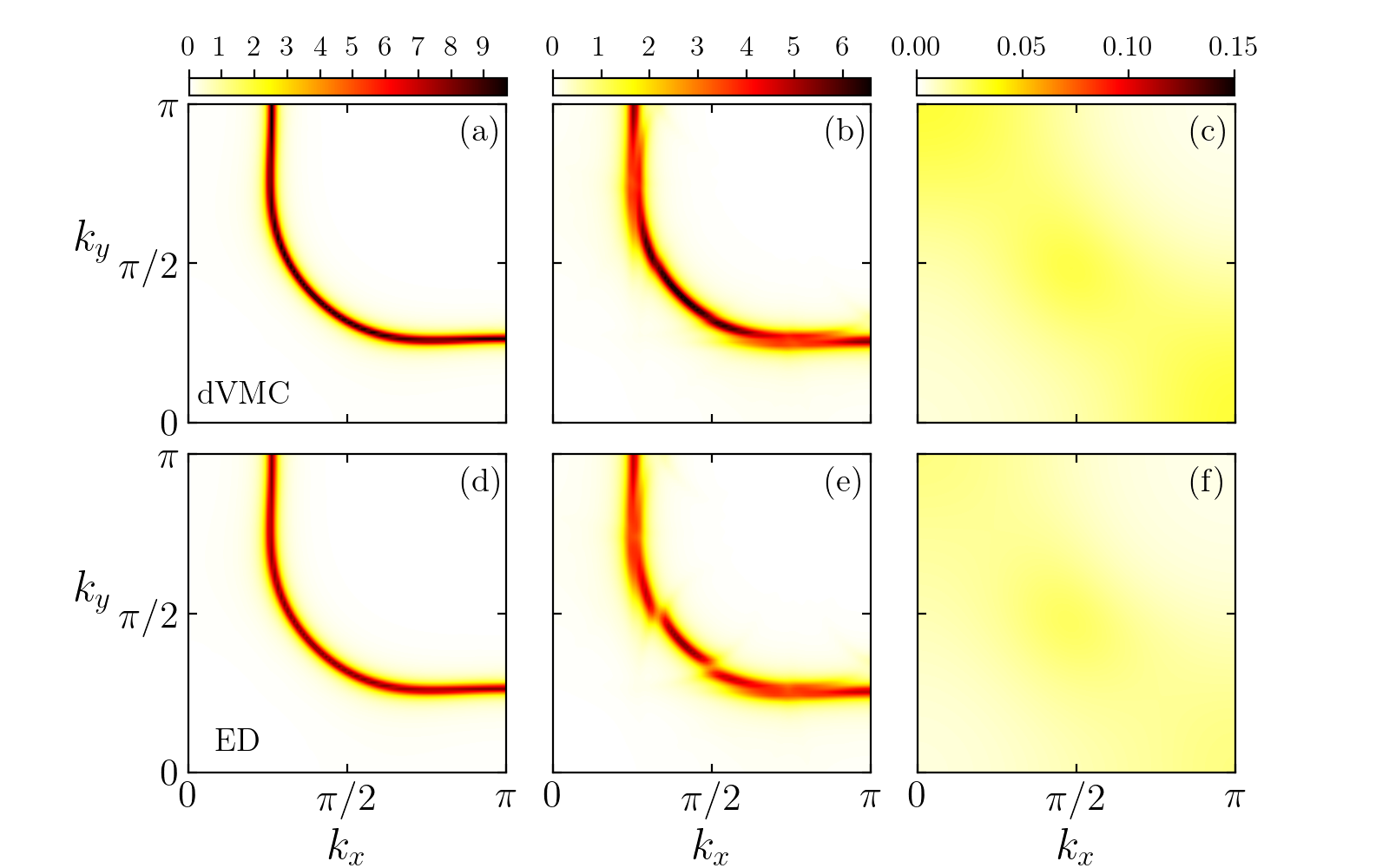}
    \end{center}
    \caption{ Fermi surface versus interaction strength, comparison with ED impurity solver for $4\times 4$ cluster at half filling with $(t,t^\prime,t^{\prime\prime})=(-1.0,0.3,-0.2)$. The top row shows the FS computed using CPT with dVMC as the impurity solver and the bottom row with ED as the impurity solver. (a),(d) $U=1$, (b),(e) $U=4$, and (c),(f) $U=8$.
     \label{fig:FS_vs_U_4x4}}
\end{figure}

\begin{figure}[!ht]
    \begin{center}
           \includegraphics[width=1.\columnwidth]{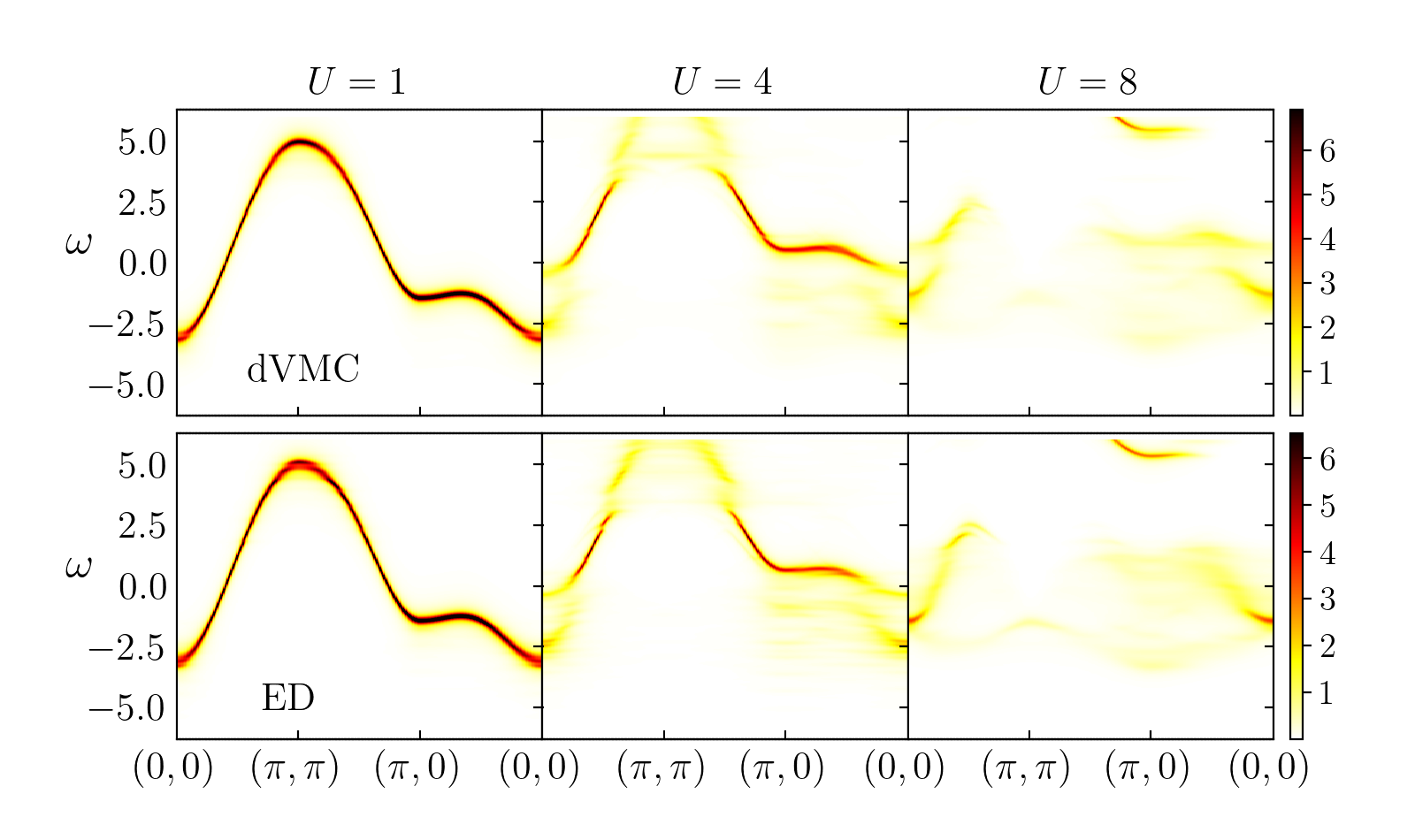}
    \end{center}
    \caption{ Spectral function versus interaction strength, comparison with ED impurity solver. The top row shows the CPT spectral function computed using dVMC as the impurity solver and the bottom row using ED as the impurity solver. This is the same system as in Fig.~\ref{fig:FS_vs_U_4x4}.
     \label{fig:Akw_vs_U_4x4}}
\end{figure}

We present a final comparison between dVMC and ED for a 4$\times$4 cluster at half-filling
at increasing values of interaction strength. In Fig.~\ref{fig:FS_vs_U_4x4} we show the Fermi
surface from dVMC and ED. 
The agreement is excellent for low to high interaction. Small features cause by the superlattice structure at $U=4$ present in ED are absent in dVMC.
Fig.~\ref{fig:Akw_vs_U_4x4} shows the spectral function along
a triangular path in the first quadrant of the BZ. Again we find excellent agreement with ED, with only
minor quantitative differences evident even at large interaction strength. The parameters of Fig.~\ref{fig:FS_vs_U_4x4} and~\ref{fig:Akw_vs_U_4x4} correspond to the ones used in the main text of the article. This is the largest square cluster that can be benchmarked against ED and from these results we conclude that CPT with dVMC is a reliable method to capture the physic observed in CPT with ED.

%...............................................................................
%...............................................................................

%...............................................................................
%...............................................................................

%===============================================================================
\section{Excitation scheme}
\label{sec:AppendixB}

In the following, we detail how the set of excited states is generated.
As described in section ~\ref{sec:general_dVMC},
the excited states are given by $\hat{c}^\dag_{i\sigma}\ket{\psi_{im\sigma}}$, for the $N_e+1$ sector
and $\hat{c}_{i\sigma}\ket{\psi_{im\sigma}}$ for the $N_e-1$ sector, where the state $\ket{\psi_{im\sigma}}$ is defined by $\hat{B}_{im\sigma}\ket{\Omega}$.
The operator $\hat{B}_{im\sigma}$ creates three distinct types of excitations.  In principle, additional types
of excitations can be used to construct this basis, but we observe that accurate results
can be obtained using only the three types included here. The first two $\hat{B}_{im\sigma}$ are the identity (corresponding to $m=0$) 
and $\hat{n}_{i\bar{\sigma}}$ ($m=1$). These two excitations are important because of their contribution to the effective basis. For $m>1$, we have another type of excitation created by the operator 
$\hat{B}_{im\sigma} = \hat{n}_{b_{im},\bar{\sigma}}\hat{n}_{b^\prime_{im},\sigma}$.
Typically it is not computationally feasible to include all possible combinations of $b_{im}$ and $b^{\prime}_{im}$ in the set of excitations. Instead,
we must select a subset of excitations, which we choose based on an empirical rule designed to identify relevant excitations. 

\definecolor{darkred}{rgb}{0.75,0.15,0.15}
\definecolor{darkblue}{rgb}{0.35,0.35,0.7}
\definecolor{darkgreen}{rgb}{0.2,0.6,0.2}

%...............................................................................
\begin{table}[]
\begin{tabular}{c|cccc}
\hline
\hline
\multicolumn{1}{l|}{$m \smallsetminus  i$} 
& $\;\;\;\;\;\; 0 \;\;\;\;\;\;$                                                       & $\;\;\;\;\;\; 1 \;\;\;\;\;\;$                                                       & $\;\;\;\;\;\; 5 \;\;\;\;\;\;$                                                       & ... 
\\ \hline
$0$                          & $\mathbb{1}$                      & $\mathbb{1}$                                   & $\mathbb{1}$                       & ... \\
$1$                          & $\color{darkred}{\hat{n}_{0\downarrow}}$                      & $\hat{n}_{1\downarrow}$                                   & $\color{darkred}{\hat{n}_{5\downarrow}}$                       & ... \\
$2$                          & $\color{darkblue}{\hat{n}_{4\uparrow}\hat{n}_{4\downarrow}}$  & $\hat{n}_{5\uparrow}\hat{n}_{5\downarrow}$                & $\hat{n}_{9\uparrow}\hat{n}_{9\downarrow}$                 & ... \\
$3$                          & $\hat{n}_{4\uparrow}\hat{n}_{1\downarrow}$                & $\hat{n}_{5\uparrow}\hat{n}_{2\downarrow}$                & $\hat{n}_{9\uparrow}\hat{n}_{6\downarrow}$                 & ... \\
...                          & ...                                                       & ...                                                       & ...                                                        & ... \\ 
$9$                          & $\hat{n}_{1\uparrow}\hat{n}_{1\downarrow}$                & $\hat{n}_{2\uparrow}\hat{n}_{2\downarrow}$                & $\color{darkblue}{\hat{n}_{6\uparrow}\hat{n}_{6\downarrow}}$   & ... \\
$10$                         & $\hat{n}_{1\uparrow}\hat{n}_{2\downarrow}$                & $\color{darkred}{\hat{n}_{2\uparrow}\hat{n}_{6\downarrow}}$   & $\hat{n}_{6\uparrow}\hat{n}_{13\downarrow}$                & ... \\
$11$                         & $\hat{n}_{8\uparrow}\hat{n}_{8\downarrow}$                & $\color{darkblue}{\hat{n}_{0\uparrow}\hat{n}_{0\downarrow}}$  & $\hat{n}_{4\uparrow}\hat{n}_{4\downarrow}$                 & ... \\
...                          & ...                                                       & ...                                                       & ...                                                        & ... \\ 
$29$                         & $\hat{n}_{5\uparrow}\hat{n}_{5\downarrow}$                & $\color{darkgreen}{\hat{n}_{9\uparrow}\hat{n}_{9\downarrow}}$ & $\hat{n}_{1\uparrow}\hat{n}_{1\downarrow}$                 & ... \\
$30$                         & $\color{darkgreen}{\hat{n}_{5\uparrow}\hat{n}_{2\downarrow}}$ & $\hat{n}_{9\uparrow}\hat{n}_{6\downarrow}$                & $\hat{n}_{1\uparrow}\hat{n}_{13\downarrow}$                & ... \\
$31$                         & $\hat{n}_{2\uparrow}\hat{n}_{4\downarrow}$                & $\hat{n}_{6\uparrow}\hat{n}_{5\downarrow}$                & $\color{darkgreen}{\hat{n}_{13\uparrow}\hat{n}_{10\downarrow}}$ & ... \\ 
\hline
\hline
\end{tabular}
	\caption{List of $\hat{B}_{im\uparrow}$ (see Eq.~(\ref{eq:exc_states2}-\ref{eq:exc_states})), for corner ($i=0$), edge ($i=1$) and interior ($i=5$) sites. We list only a subset of $i$ and $m$ in order to remain concise. The colors correspond to the examples illustrated in Fig.~\ref{fig:exc_scheme}.}
	\label{tab:exc_list}
\end{table}
%...............................................................................

%...............................................................................
\begin{figure*}[]
    \begin{center}
           \includegraphics[width=1.\textwidth]{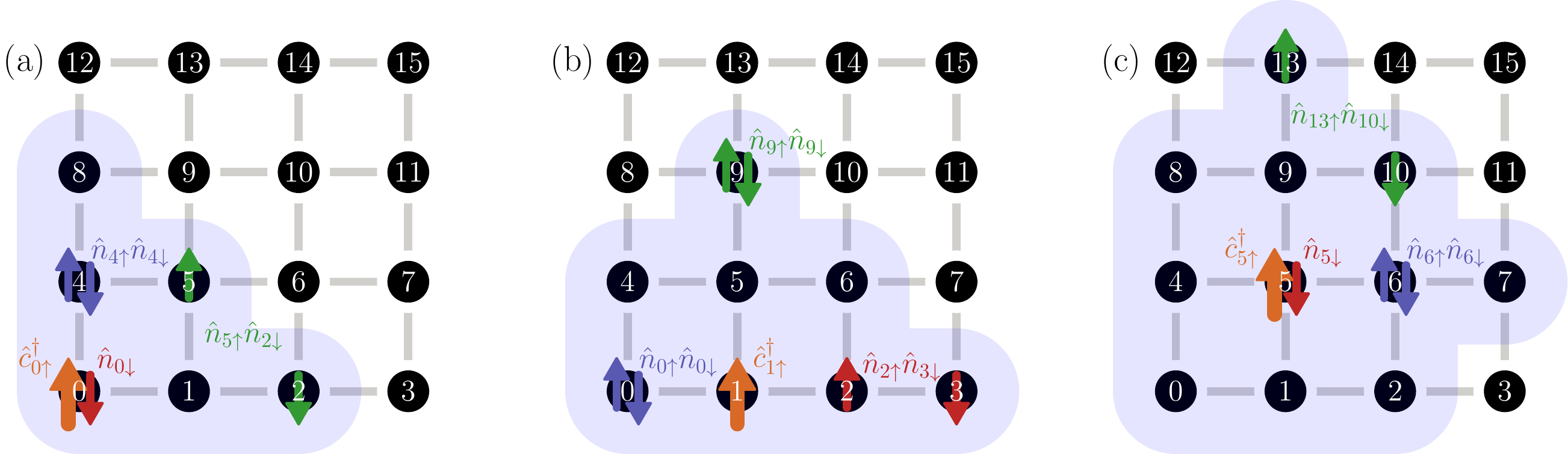}
    \end{center}
    \caption{Illustration of few $\hat{B}_{im\uparrow}$ (see Eq.~\eqref{eq:exc_states}), for (a) corner ($i=0$), (b) edge ($i=1$) and (c) interior sites ($i=5$) of  a $4\times 4$ cluster. The blue shaded region represents all the sites that can be reached from site $i$ with $n_h=2$ nearest-neighbor hops. The colored arrows (green, blue and red) represent the position and the spin of the $\hat{n}_{i,\sigma}$ operator and the orange arrow is the position and the spin of the $\hat{c}^\dag_{i\sigma}$ operator. The colors correspond to the colored entries of  table~\ref{tab:exc_list}.}
     \label{fig:exc_scheme}
\end{figure*}
%...............................................................................

As a concrete example of the procedure for selecting the $\hat{B}_{im\sigma}$, let us consider the case of a $4\times 4$ cluster. This cluster is illustrated in Fig.~\ref{fig:exc_scheme}.
The set of sites within $n_h=2$ hops for a 4$\times$4 cluster is indicated by the blue shaded region, first
for a corner site ($i=0$), followed by an edge site ($i=1$), and then an interior site ($i=5$). Only $3$ values of $m$ are illustrated in each of these 3 cases (3 values of $i$) but many more excitations are possible. Here, we used $N_{\rm exc}=32$. In table~\ref{tab:exc_list}, we show a partial list of all the excitations we keep for this cluster, with $n_h=2$, including those shown in Fig.~\ref{fig:exc_scheme}. 

We note that the number of neighbors within $n_h=2$ hops (blue shaded region) is different depending on the site $i$. The corner site ($i=0$) has only $N^{\rm min}_b=5$ while an interior site ($i=5$) has $N^{max}_b=10$. As a computational convenience we choose an equal number of excitations for each cluster site, but in principle the number of excitations can be different for each site. 
The maximum number of excitations per site (assuming we require all sites to have the same number of excitations) is given by
$N_{\rm exc} = 2 + N^{\rm min}_b(N^{\rm min}_b + 1)$, where $N^{\rm min}_b$ is the number of neighbors for the site $i$
with the fewest neighbors, generally a corner site. For example, in Fig.~\ref{fig:exc_scheme}(a), $N^{\rm min}_b=5$ thus $N^{\rm exc}=32$. The $N^{\rm min}_b$ and $(N^{\rm min}_b + 1)$ factors come from the number of positions in the blue shaded region where we can put an electron of same spin $\sigma$ and different spin $\bar \sigma$ respectively, for excitations associated to Eq.~\eqref{eq:exc_states}.
For non-corner sites there are more neighboring sites and therefore more possible excitations, but we choose to truncate the set of excitations for these sites in order to keep $N_{\rm exc}=32$ for all $i$, as listed in table~\ref{tab:exc_list}.

Finally, in table~\ref{tab:exc_number} we show the number of excitations kept versus $n_h$ for two different clusters: $4 \times 4$ and $8 \times 8$. This table shows the growth of $N_{\rm exc}$ as a function of $n_h$. 
The total number of excitations, $N_{\rm exc}N$, sets the number of rows and columns, i.e. the dimension, of the matrices $\mathbf{S}$, $\mathbf{M}$, $\mathbf{U}$ and $\mathbf{E}$ from Eqs.~\eqref{eq:Gpm1} to \eqref{eq:Gpm4}, and also roughly gives the number of eigenenergies and poles contained in the Green function.
If this number grows too large it becomes too computationally expensive to obtain the eigenvalues of $\mathbf{S}$ and $\mathbf{M}$. This threshold is indicated by the numbers in red in table~\ref{tab:exc_number}. As a point of reference, for a $2\times 2$ cluster, this scheme yields a maximum number of excitations per site $N_{\rm exc} = 14$, which is relatively small and therefore easily computable. 

%...............................................................................
\begin{table}[h]
\begin{tabular}{c|ccc|ccc}
\hline
\multicolumn{1}{l|}{} & \multicolumn{3}{c|}{$4 \times 4$}           & \multicolumn{3}{c}{$8 \times 8$}            \\ \hline
$n_h$                 & $\; N^{\rm min}_b \;$ & $\; N_{\rm exc} \;$ & $\; N_{\rm exc}N \;$ & $\; N^{\rm min}_b \;$ & $\; N_{\rm exc} \;$ & $\; N_{\rm exc}N \;$ \\ \hline
1                     & 2                & 8         & 128          & 2                & 8         & 512          \\
2                     & 5                & 32        & 512          & 5                & 32        & 2048         \\
3                     & 9                & 92        & 1472         & 9                & 92        & 5888         \\
4                     & 12               & 158       & 2528         & 14               & 212       & $\color{darkred}{13568}$          \\
5                     & 14               & 212       & 3392         & 20               & 422       & $\color{darkred}{27008}$        \\
6                     & 15               & 242       & 3872         & 27               & 758       & $\color{darkred}{48512}$        \\ \hline
\end{tabular}
	\caption{Number of excitations $N_{\rm exc} = 2 + N^{\rm min}_b(N^{\rm min}_b + 1)$ and its relation to the number of neighbor of the corner site $N^{\rm min}_b$ as a function of the number of hops $n_h$. We show these numbers for two square clusters: $4 \times 4$ and $8 \times 8$. $N_{\rm exc}N$, the size of the matrices, is also shown.}
	\label{tab:exc_number}
\end{table}
%...............................................................................

For the sake of clarity and simplicity, we have not included all the details of our method for generating excitations with the operator $\hat{B}_{im\sigma}$.
Generally, we find that $\hat{B}_{i m\sigma}$ where the sites $b^{(\prime)}_m$ are near the site $i$
tend to be more important. In any case, provided the number of electrons $N_e$ and the spin $S_z$ are conserved, any excitation generated by
the operator $\hat{B}_{im\sigma}$ can contribute to the sampling of the effective basis. For sufficiently large sets of excitations generated according to these principles, our results for the converged Green function 
are relatively insensitive to the specific excitations chosen.

%\bibliography{outline,dvmc}
%

\end{document}